\def\spose#1{\hbox to 0pt{#1\hss}} %
\def\Dt{\spose{\raise 1.5ex\hbox{\hskip3pt$\mathchar"201$}}}    
\def\dt{\spose{\raise 1.0ex\hbox{\hskip2pt$\mathchar"201$}}}    

\def\lta{\mathrel{\spose{\lower 3pt\hbox{$\mathchar"218$}}
     \raise 2.0pt\hbox{$\mathchar"13C$}}}

\def\gta{\mathrel{\spose{\lower 3pt\hbox{$\mathchar"218$}}
     \raise 2.0pt\hbox{$\mathchar"13E$}}}

\documentstyle[12pt,psfig]{article}
\begin{document}

\vspace*{1.0in}

\begin{center}
				Viscous liquid flow on Martian dune slopes

\vspace*{0.5in}
				Anthony R. Dobrovolskis 

				SETI Institute

				245-3 NASA Ames Research Center 

				Moffett Field, CA 94035-1000 

				Email: anthony.r.dobrovolskis@nasa.gov 

\vspace*{0.5in}
					2014 May 27

\vspace*{1.0in}
					ABSTRACT
\end{center}

The observed temporary dark streaks on some dune slopes on Mars may be 
due to thin sheets of water (or some other liquid) trickling downhill. 
This note corrects conceptual errors in a previous paper 
(M\"{o}hlmann and Kereszturi 2010, {\it Icarus} {\bf 207}, 654--658) 
which affect the velocity profile of such flows, 
and produce over-estimates of their depths 
and mass fluxes by factors of almost two.

\newpage

\section{INTRODUCTION}

Dark streaks have been observed propagating downhill 
on high-latitude dunes on Mars during local springtime. 
M\"{o}hlmann and Kereszturi (2010; hereafter MK2010) 
have attributed these streaks to the flow 
of thin sheets of water (or some other liquid), 
and derived a relation between the measured speed 
of such flows and the thickness of the liquid layer. 

The purpose of this note is to correct two conceptual errors in MK2010 
which affect the derivation of the velocity profile of such flows, 
and interpretations of their observed speed, depth, and mass flux. 
(Note also that the last phrase of Section 2 of MK2010 should state 
that direct measurements of local increases in the surface temperature  
due to dune darkening are {\it not} available yet.)

\section{MODEL}

My model is fundamentally the same as that in MK2010 
({\it cf.} their Fig. 2): 
a sheet of liquid with uniform thickness $h$, 
constant density $\rho$, and dynamic viscosity $\eta$,  
is flowing down an inclined plane at a fixed angle $\alpha$ 
from the local horizontal, under the influence 
of the vertical acceleration of gravity $g$. 

As in MK2010, let $x$ be the downslope coordinate, 
$z$ the upward coordinate perpendicular to $x$, 
and $y \equiv z/h$. Henceforth $x$ can be ignored, 
and the speed of the flow can be written simply as 
$v(z)$ or $v(y)$. 

However, MK2010 make their first conceptual error 
in using the Navier-Stokes equation for incompressible 
flow of a fluid with constant viscosity; their Eq. (2) is 
\begin{equation}
		\eta \frac{d^2}{dz^2} v(z) = -\rho g \sin\alpha. 
\end{equation}

Most liquids may safely be treated as incompressible, 
but MK2010 next assume that the viscosity of the fluid 
depends very strongly on the vertical coordinate, 
due to freezing at its upper and lower surfaces; 
their Eq. (3) is 
\begin{equation}
		\eta(y) = \frac{\eta_0}{y[1-y]}. 
\end{equation}

Under these circumstances the usual Navier-Stokes equation 
(1) is inadequate, and a more general version is required: 
\begin{equation}
		\frac{d}{dz} [ \eta \frac{d}{dz} v(z) ] = 
		\frac{d}{dy} [ \eta \frac{d}{dy} v(y) ]/h^2 
                        = -\rho g \sin\alpha. 
\end{equation}

Integrating Eq. (3) above gives the viscous stress 
\begin{equation}
		\eta \frac{d}{dy} v(y) = C -y h^2 \rho g \sin\alpha 
\end{equation}
as a linear function of height $y$. 
Then equating this to zero at the top $y=1$ gives 
\begin{equation}
			C = h^2 \rho g \sin\alpha 
\end{equation}
by the no-stress boundary condition at the free surface. 

Now by using Eq. (2), Eq. (4) can be rearranged as 
\begin{equation}
		\eta_0 \frac{d}{dy} v(y) = C y[1-y]^2 = C [y -2y^2 +y^3], 
\end{equation}
and integrated as 
\begin{equation}
		v(y) = C[y^2/2 -2y^3/3 +y^4/4]/\eta_0 +C'. 
\end{equation}
Here the second constant of integration $C'$ must vanish 
by the no-slip boundary condition $v(0)=0$ at the bottom. 

As their second conceptual error, MK2010 integrate Eq. (1) twice, 
but then apply a symmetry condition $dv/dy = 0$ at the mid-plane $y=1/2$. 
This is equivalent to imposing a no-slip condition at the free surface, 
rather than the correct no-stress condition. 
As a result, MK2010 obtain the spurious solution 
\begin{equation}
                v(y) = C[y/12 -y^3/6 +y^4/12]/\eta_0 ; 
\end{equation}
note also that their Eq. (5) has the opposite sign as Eq. (8) above.

\begin{figure}
\centerline{\psfig{figure=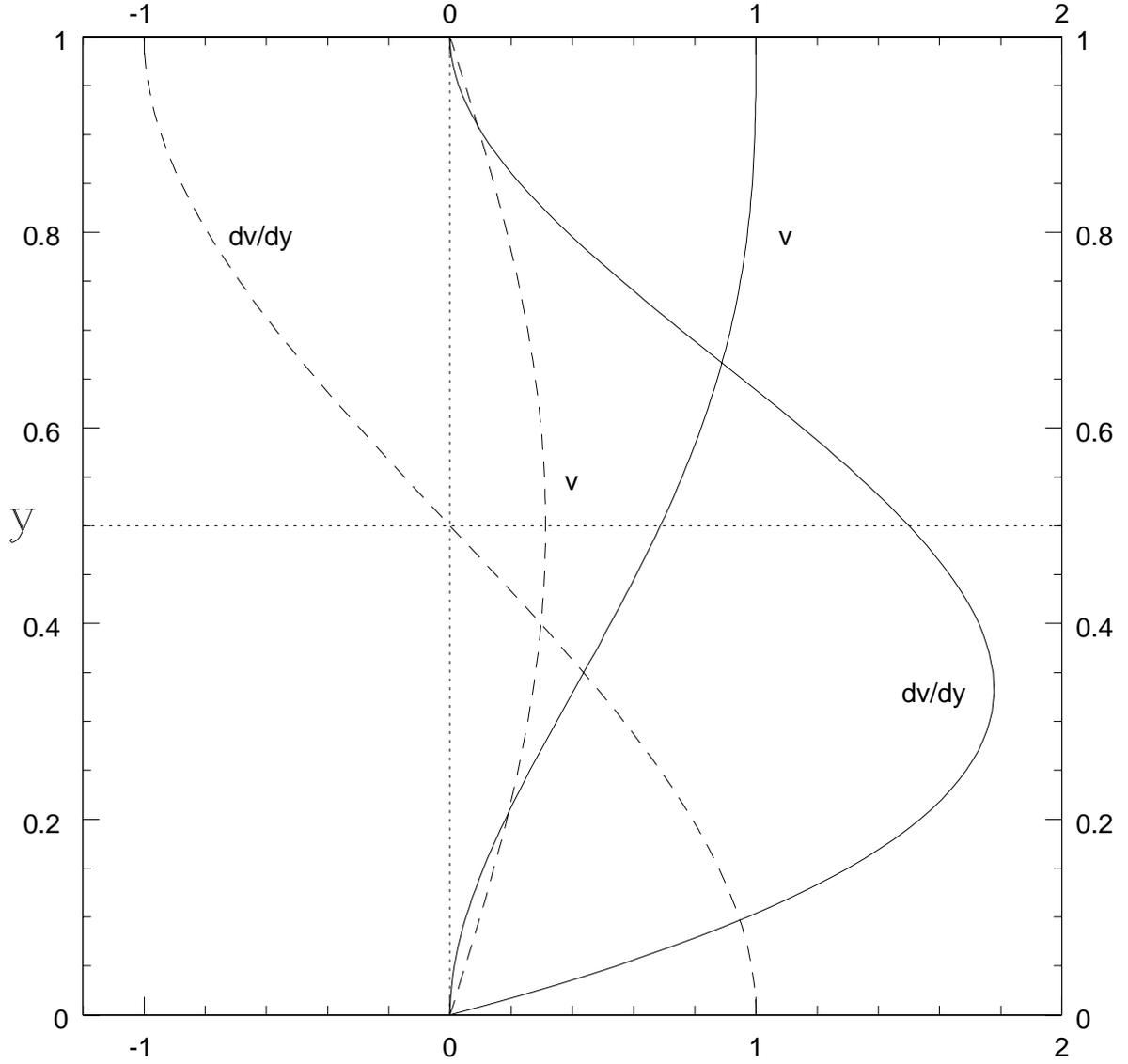,width=6.5in,height=6.5in}}
\caption{ Profiles of flow velocity $v$ and shear $dv/dy$, 
both normalized by $\frac{C}{12 \eta_0}$, 
as functions of dimensionless height $y$. 
Dashed curves refer to solution (8) after MH2010, 
while solid curves refer to my corrected solution (7). 
The horizontal dotted line denotes the midplane $y=1/2$, 
while the vertical one denotes $v=0$ or $dv/dy=0$. }
\end{figure}

\newpage

\section{RESULTS}

Figure 1 compares solution (8) above (dashed curves) 
with my corrected solution (7) (solid curves). Note how $v$ 
is symmetric with respect to the midplane $y=1/2$ for solution (8), 
while its corresponding shear $dv/dy$ is antisymmetric. 
The viscous stress $\eta \; dv/dy$ is not shown, 
because it is infinite at $y=0$ and $y=1$ for solution (8). 

Solution (8) may be rewritten in symmetric form as 
\begin{equation}
                v(\zeta) = C[5/192 -\zeta^2/8 +\zeta^4/12]/\eta_0 , 
\end{equation}
where $\zeta \equiv y -1/2$. Eq. (9) above makes it clear 
that this velocity profile is not a parabolic curve, 
as in channel flow, but rather a quartic curve. In fact, 
it is a biquadratic; that is, a quadratic in $\zeta^2$. 

The flow speed $v$ vanishes at both the top and bottom for solution (8),
and peaks at $v_{max} = \frac{5 C}{192 \eta_0}$ at the midplane $y=1/2$ ($\zeta=0$);
while the corresponding shear $dv/dy$ ranges from $\frac{C}{12 \eta_0}$
at the bottom, through zero at the midplane, to $\frac{-C}{12 \eta_0}$ at the top.

In contrast, neither $v$ nor $dv/dy$ possesses any symmetry 
for my corrected solution (7). Note how $v$ vanishes at the bottom $y=0$, 
and peaks at $v_{max} = \frac{C}{12 \eta_0}$ at the top $y=1$; 
while the shear $dv/dy$ vanishes at both the top and bottom, 
and peaks at $\frac{4C}{27 \eta_0}$ at $y=1/3$. 

Note that the peak speed $v_{max}$ is 16/5 = 3.2 times greater for my solution (7) than 
for solution (8). MK2010 also assumed that $v_{max}$ is the observed propagation speed of 
the dark dune streaks, and used it to find their Eq. (7) for the thickness $h$ of the flow: 
\begin{equation}
		h \approx \sqrt{ \frac{192 \eta_0 v_{max}}{5 \rho g \sin\alpha} }. 
\end{equation}

Using my solution (7) instead to estimate $h$ gives 
\begin{equation}
                h \approx \sqrt{ \frac{12 \eta_0 v_{max}}{\rho g \sin\alpha} }. 
\end{equation}
Note that $h$ from formula (11) above is only $\sqrt{5/16} \approx$ 0.559 times as deep 
as from formula (10); for the example given by MK2010, 
Eq. (11) gives a layer of brine only 1.2 mm thick, rather than 2.2 mm from Eq. (10). 

Furthermore, the mean speed of the flow may be defined as 
\begin{equation}
		\bar{v} \equiv \int_0^1 v(y) dy . 
\end{equation}
Then the mean speed for my Eq. (7) is $\frac{C}{20 \eta_0}$, or 0.60 $v_{max}$. 
For comparison, the mean speed for Eq. (8) is only $\frac{C}{60 \eta_0}$, 
or 0.64 $v_{max}$. In either case, note that the flux (mass per unit length 
per unit time) is just $\rho h \bar{v}$. Then for a given $v_{max}$, 
the flux from my Eq. (7) is only 0.524 times as great as from Eq. (8).

\newpage

\section{DISCUSSION}

MK2010 have attributed dark streaks observed propagating downhill 
on high-latitude dunes on Mars to the flow of thin sheets of water 
(or some other liquid), and derived relation (10) between the 
measured speed of such flows and the thickness of the liquid layer. 
However, MK2010 made two important conceptual errors in their derivation: 

First, they used the Navier-Stokes equation for flow of an incompressible 
fulid of constant viscosity; most liquids are nearly incompressible, 
but MK2010 also assumed a very non-uniform viscosity, 
rendering the Navier-Stokes equation inapplicable. 

Second, they assumed that the velocity profile of the flow is symmetric about its midplane; 
this is equivalent to imposing a no-slip boundary condition at the top of the liquid layer. 
A no-slip condition is appropriate at the bottom of the flow, 
but there is no justification for it at the top; 
rather, a no-stress condition is required there. 

Correcting both errors results in the revised relation (11) 
between the speed and thickness of the flow. 
Comparing Eqs. (10) and (11) then reveals that Eq. (10) 
over-estimates the thickness of the liquid layer 
and the corresponding mass flux by factors of almost two. 

However, the viscosity of some concentrated brines on the surface of Mars 
may vary with composition by several orders of magnitude ({\it cf.} Fig. 5 
of MK2010). Because the derived depth of the liquid layer is inversely 
proportional to the square root of the viscosity in both Eqs. (10) and (11), 
the resulting uncertainty in thickness may be dominated 
by uncertainty in composition, rather than by errors in the formula.

\setlength{\parindent}{-0.2in}

\begin{center}
                                REFERENCE
\end{center}

M\"{o}hlmann, D., and A. Kereszturi (2010). 
Viscous liquid film flow on dune slopes of Mars. 
\\ {\it Icarus} {\bf 207}, 654--658.

\end{document}